\documentstyle[aas2pp4]{article} 
\def\vecn{\mbox{\bf n}}
\def\vecS{\mbox{\bf S}}
\def\vecI{\mbox{\bf I}}
\def\vecw{\mbox{\bf w}}

\tightenlines

\begin{document}

\title{Optimal Masks for Low-Degree Solar Acoustic Modes}
\author{T. Toutain}
\affil{D\'epartement CASSINI, Observatoire de la C\^ote d'Azur, Nice, France}
\authoremail{toutain@obs-nice.fr}
\author{ A. G. Kosovichev}
\affil{W.W.Hansen Experimental Physics Laboratory,\\
Stanford University, Stanford, CA 94305-4085}
\authoremail{akosovichev@solar.stanford.edu}

\begin{abstract}
We suggest a solution to an important problem of observational helioseismology
of the separation of lines of solar acoustic (p) modes of low angular degree
in oscillation power spectra by constructing optimal masks for Doppler
images of the Sun.
Accurate measurements of oscillation frequencies of low-degree modes
are essential for the determination of the structure and rotation of the solar
core. However, these measurements for a particular mode
are often affected by leakage of
other p modes arising when the Doppler images are projected on to
spherical-harmonics masks. The leakage results in overlaping peaks
corresponding to different oscillation modes in the power spectra.
In this paper we present a method for calculating
optimal masks for a given (target) mode by
minimizing the signals of other modes appearing in its vicinity.
We apply this method to
time series of 2 years obtained from Michelson Doppler Imager (MDI) instrument
on board SOHO space mission and demonstrate its ability to
reduce efficiently the mode leakage.
\end{abstract}


\section{Introduction}
Accurate measurements of frequencies and other physical properties of solar
oscillations are crucial for understanding the internal structure and
dynamics of the Sun.
Up to now the determination of low-degree p-mode parameters like cyclic
frequency
$\nu$, linewidth
or frequency splitting is based on fitting the so-called $m$-$\nu$
diagrams obtained by
projecting time series of solar images (usually Dopplergrams)
on to a basis of spherical harmonics
of angular degree $l$ and azimuthal order $m$. Because
only one hemisphere of the Sun is observed the projection on to
spherical harmonics is not orthogonal. This results in
leakage of modes which makes the line fitting quite complicated especially when
modes overlap with the target mode. We have therefore decided to apply a method
originally developed by Kosovichev (1986) based on a singular value
decomposition
(SVD) technique in order to minimize the leakage.

We discuss in  Sec. 2 the
principle of the method, and introduce the concepts of `global' and `local'
optimal masks. The global optimal masks are designed to minimize the
contribution
of all modes except the target one. The local masks minimize only modes within
a narrow frequency interval centered at the target mode.
In Sec. 3 we show the theoretical efficiency of
the method compared to the usual spherical harmonics projection.  In Sec. 4
we illustrate this method with a 2-year time
series of MDI LOI-proxy data, obtained by rebinning of the original 1024 by
1024 CCD
pixels
into 180 bins (Hoeksema et al., 1998; Scherrer et al., 1995)
according to the shape of the LOI instrument (Appourchaux et al. 1997).

\begin{figure}[t]
\epsscale{1}
\plotone{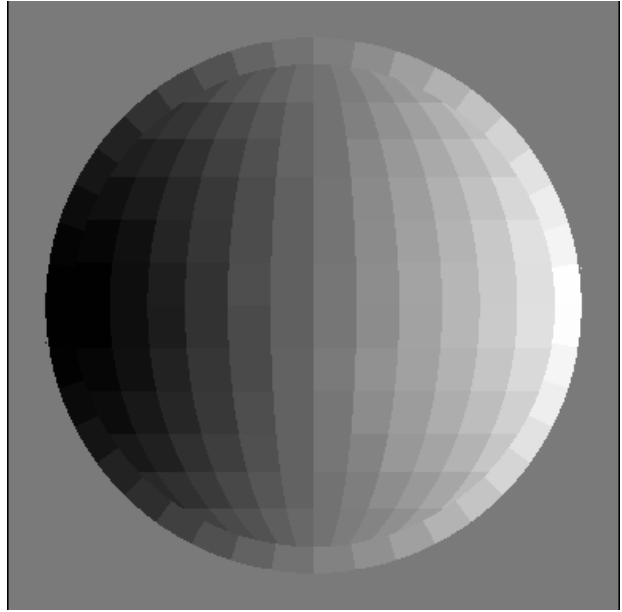}
\caption{An example of MDI LOI-proxy Dopplergrams. The full-disk
1024x1024-pixel
MDI Dopplergrams are rebinned on board SOHO into 180 bins with the boundaries
along
heliospheric meridians and parallels. The variation from dark to light colors
is
due to solar rotation.\label{figure1}}
\end{figure}

\section {The optimal-mask method}

The aim of the optimal masks (Gough, 1984 and Christensen-Dalsgaard, 1984) is
to find
an optimal linear combination of $N$ signals arising from $N$ bins
of the solar image in order to produce a
time series containing mainly the contribution of a mode with given
 $l$ and $m$,
the so-called target mode, and minimizing the contribution of others
modes.
It is particularly important to minimize the contribution of modes
which are close to the target mode in the power spectrum. These are mainly
modes
of the same rotationally split multiplet as the target mode. At frequencies
about 2.5 mHz and higher the linewidth of p modes becomes larger than
the separation between the peaks. Therefore, it is important to separate
the overlapping modal lines. We describe this method in the application to
the MDI LOI-proxy data. However, this method is quite general and can be
applied
to other spatially resolved helioseismic data.

Because we observe only one hemisphere of the
Sun, the commonly used spherical-harmonic masks are not able to remove all the
components of the target mode multiplet.  Instead
of using the spherical harmonics masks we apply a
SVD method to derive the optimal masks.  To apply this
method we first need to model the signal produced in each
image bin by a mode of given angular degree $l$,
azimuthal order $m$ and radial order $n$.
For simplicity, in this paper, the horizontal component of velocity is
neglected because we are only interested in low-degree p modes.
The inclusion of the horizontal component is straightforward, and generally
improves the quality of the optimal masks in the low-frequency part of
the solar oscillation spectrum where this component becomes significant.
Hence, to the first
order, the signal of a normal mode in each pixel $j$ of the CCD detector is:
\begin{equation}
v^{(j)}_{nlm}(t) = \Re\left\{V_{nlm}e^{i\omega_{nlm}t}s^{(j)}_{lm}(t)\right\},
\end{equation}
where $V_{nlm}$ is the mode amplitude, $\omega_{nlm}$ is the mode frequency,
and
$s^{(j)}_{lm}(t) = Y_{lm}[\theta_{j}^{*}(t),\phi_{j}^{*}(t)]\mu_{j}$
is the sensitivity function which depends on the spatial eigenfunction,
spherical harmonic
$Y_{lm}$, and the factor $\mu_{j}$ arising from the projection of the
radial velocity onto the line of sight. Both $Y_{lm}$ and $\mu_{j}$
are expressed in terms of the
heliographic coordinates,
$\theta_{j}^{*}(t)$ and $\phi_{j}^{*}(t)$, of the pixels, which are generally
functions
of time $t$.   We have to take into account the fact that the
orientation of the solar image,
which is usually described by two angles of the rotation axes
with respect to the ecliptics, $P_0$ and $B_0$,
and the image size on the detector are possibly changing with time.
The time dependences of the $P_0$ and $B_0$ angles should therefore be taken
into
account by rotating the spherical coordinates accordingly.
In our case the variations of the image size were not taken into account
because
the data were combined into fixed 180 bins on board spacecraft prior to any
mask
application. An example of the rebinned Dopplergrams (`MDI LOI-proxy') is shown
in Fig. 1.

 The k$^{th}$
signal of each of the $N$ bins is obtained by averaging the signals coming from
$N_{k}$ CCD pixels withing this bin
as defined by the mapping between the MDI pixels and the LOI-proxy bins:
\begin{equation}
v^{(k)}_{nlm}(t) = \Re\left\{V_{nlm}e^{i\omega_{nlm}t}S^{(k)}_{lm}(t)\right\},
\end{equation}
where
$S^{(k)}_{lm}(t) = \frac{1}{N_{k}}\sum_{j=1}^{N_{k}} s^{(j)}_{lm}(t)$
is  $S^{(k)}_{lm}$ is the sensitivity of bin $k$ to a normal mode of angular
degree $l$
and order $m$.
To extract the signal of a target mode of $n_0$, $l_0$ and $m_0$
we combine the $N$ signals with $N$ weights
$w_{n_0l_0m_0}^{(1)}$,...,$w_{n_0l_0m_0}^{(N)}$:
\begin{eqnarray}
v_{n_0l_0m_0}(t) = \sum_{k=1}^N  w_{n_0l_0m_0}^{(k)}
v^{(k)}_{nlm}(t)=\nonumber\\
\Re\left\{V_{nlm}e^{i\omega_{nlm}t}\left[\sum_{k=1}^N
w_{n_0l_0m_0}^{(k)}S^{(k)}_{lm}(t)\right]\right\}.
\end{eqnarray}
Ideally, if  the sum in the square brackets is equal to
 the product of two Kroneker
symbols, $\delta_{ll_{0}}\delta_{mm_{0}}$, then $v_{n_0l_0m_0}(t)$
consists only of the signal of the target mode. However, in practice,
this sum does not represent $\delta_{ll_{0}}\delta_{mm_{0}}$
and contains not only the target mode but also contributions from other
modes.

Vector $\vecw_{n_0l_0m_0}\equiv\{w_{n_0l_0m_0}^{(1)},...,w_{n_0l_0m_0}^{(N)}\}$
defines an
optimal mask if this vector is chosen to maximize the signal for the
target mode and minimizes the signals for other, say $M$, modes
located close to the target
mode in the Fourier domain. These conditions can be written as a minimization
of the quadratic form:
\begin{equation}
r=||\vecS\cdot\vecw - \vecI||^2,
\end{equation}
where $\vecS$ is a $(M+1)\times N$ matrix with elements
$S^{(k)}_{nlm}(t)$ and $\vecI$ is a
$(M+1)$-element vector with elements $\delta_{ll_{0}}\delta_{mm_{0}}$, $l_0$
and $m_0$
being the angular degree and order of the target mode.
  
 Finding $\vecw$ is
equivalent to finding a least-squares solution to
the system of linear equations, $\vecS\cdot\vecw = \vecI$.
Depending on $M$, the number of modes we want to minimize, we have a system
which is
over-determined (N $>$ M+1) or under-determined (N $<$ M+1).
For both cases the problem is efficiently solved using the
singular value decomposition (SVD) of the matrix $\vecS$
(e.g. Press et al. 1992).
This method optimally solves the
linear system and its solution $\vecw$ gives the optimal mask.
This  mask provides an efficient filtering of the selected $M$ modes.
Of course, the minimization is less efficient when $M$ is large.
There is therefore a trade-off between the
number of modes to minimize and the efficiency of this minimization.

Matrix $\vecS$ describes the sensitivity of each detector bin to the mode of
oscillation.
For the special case of an under-determined system (N $<$ M+1)
this technique is equivalent to the technique of diagonalization of
the leakage matrix (Appourchaux et al., 1998).
The SVD method allows us to consider general cases.
With this technique we can specifically decide which modes must be filtered
and to what extend their contribution to the power spectra
should be reduced, making the technique extremely flexible.
In practice we filter as much as
possible the modes having frequencies falling in a given frequency window
around the
target making a kind of local cleaning of the spectra.  We therefore call the
corresponding masks `local optimal masks' to distinguish from the `global
optimal
masks' obtained when minimizing the contribution of all the modes not only
those around
the target mode.  The larger the window around the target the more modes have
to be
minimized,
and thus the less efficient is the filtering.
In principle we are able to remove efficiently as many
modes as the number of the binned signals, 180 in our case. Above that number
the SVD acts
as a least square minimization and the modes cannot be fully filtered.
In practice, the efficiency of the filtering depends on the selected mode set
and the noise level.

The solar noise obviously will be less filtered than modes by the masks because
its
spatial structure is incoherent.  Thus the
signal-to-noise ratio of the target mode depends on the mask.
More efficient masks typically require higher weights $w_{nlm}^{(k)}$,
and, therefore amplify the noise. This leads to the trade-off between
the mask efficiency and the noise level. To adjust this trade-off we slightly
modify Eq.~(4) and add the noise contribution through a regularization term.
Then the quadratic form to minimize is
\begin{equation}
r=||\vecS\cdot\vecw - \vecI||^2 + \alpha ||\vecn \cdot \vecw ||^2,
\end{equation}
where $\vecn=\{n^{(1)},...,n^{(N)}\}$ is the vector of the standard deviation
of the noise
in each of the $N$ bins, and $\alpha$ is the regularization parameter.
The SVD technique is still applicable here. Eq. (4)
corresponds to the case of $\alpha$=0 which means no regularization.

The noise level is estimated by taking the standard deviation of the
signal in each bin, assuming the main contributor is the solar noise.
Generally, the noise level is the lowest at the center of the solar disk and
increases
toward the limb. Table 1 gives a sample of the relative noise level
of the MDI measurements as a function of the angular distance from the
disk center. Obviously, the absolute noise level is not required for Eq.(5).
\begin{table}
\caption{Relative noise level, $n$, in the MDI data as a function of the
angular
distance, $d$, from the disk center.}
\begin{center}
\begin{tabular}{crrrrrr}
\hline\hline
$d$, deg.  & 1.72 & 11.0 & 20.5 & 30.7 & 42.1 & 56.1\nl
$n$  & 1.00 & 1.15 & 1.40 & 1.73 & 1.99 &  2.27 \nl
\hline
\end{tabular}
\end{center}
\end{table}

The regularization parameter $\alpha$ is chosen
such as the amplitudes of the filtered modes is reduced to the
level of noise.
It turns out that the
signal-to-noise ratio obtained with the local optimal masks is
similar to what is obtained
with spherical-harmonic masks, though it may be somewhat smaller (up to 50\%).
Only for $l$=0
the difference in the signal-to-noise ratio between the two masks
can reach a factor of two or more.  Because of solar noise we also avoid using
the
limb bins. These
bins improve very little  the efficiency of the masks but contribute
significantly to the noise. Removing not only limb bins but also
the next round of bins helps to increase the signal-to-noise ratio
at low frequencies for $l=0$ modes, but this is not effective for modes of
higher
degrees.

The power of this technique is also that the optimal masks for any $P_0$
and $B_0$ angles
are easily obtained applying the spherical harmonic rotation matrix to the
optimal
masks with $P_0$=0 and $B_0$=0. Therefore it is possible as for MDI to apply
the
optimal masks for each day according to the value of $B_0$ which is changing
with time
without recomputing the SVD problem which is a heavy computational task .

\section{Comparison of spherical-harmonic masks and optimal masks}
\begin{figure}
\epsscale{1.1}
\plotone{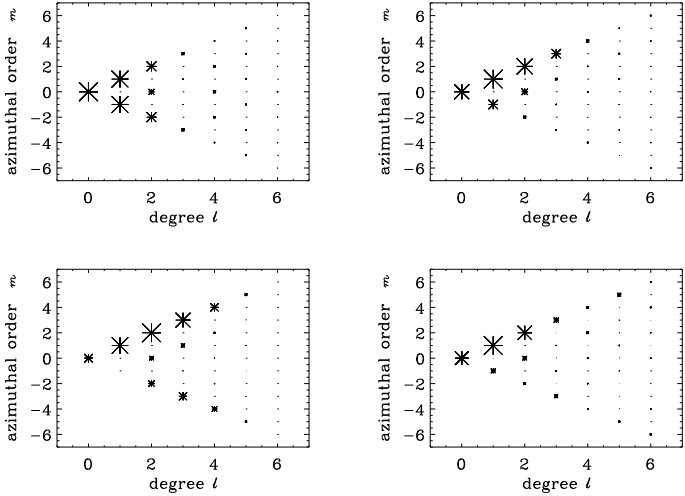}
\caption{Amplitude maps for a disk-integrated mask (top, left),
spherical-harmonics
masks(top, right),local optimal masks (bottom, left) and global optimal masks
(bottom right) with target $l$=1, $m$=1. The size of the symbols on this map is
proportional to mode amplitudes.\label{l1m1}}
\end{figure}
\begin{figure}
\epsscale{1.1}
\plotone{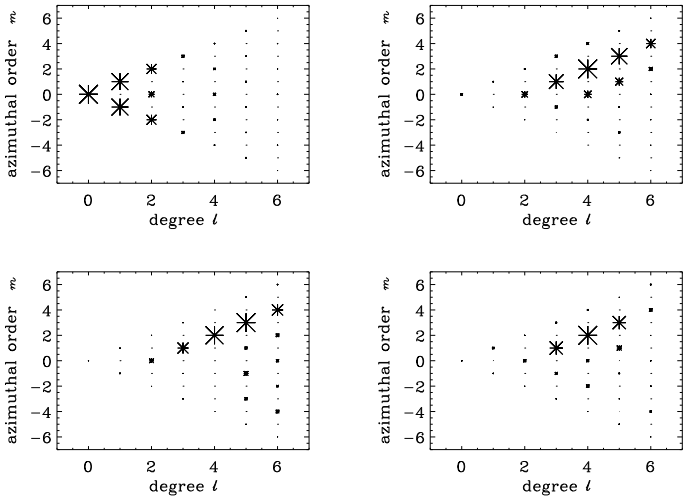}
\caption{The same as in Fig. 2 but for $l$=4,
$m$=2.\label{l4m2}}
\end{figure}

\begin{figure}
\epsscale{1.1}
\plotone{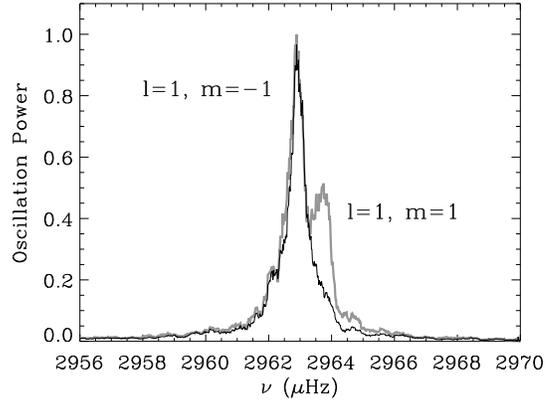}
\caption{Power spectrum for $l$=1 $m$=-1 mode using the spherical-harmonic mask
(grey curve)
and local optimal masks (black curve).\label{l0sph}}
\end{figure}
\begin{figure}
\epsscale{1.1}
\plotone{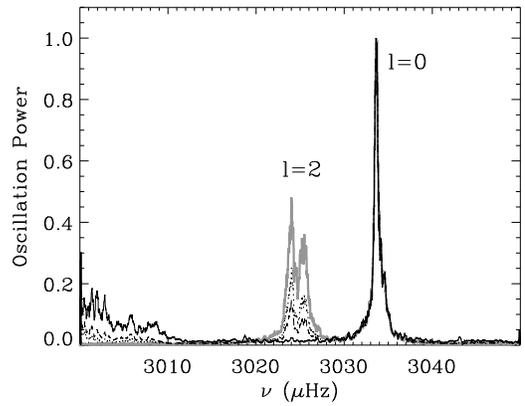}
\caption{Power spectrum around target mode $l$=0 at 3034 $\mu$Hz for
local optimal masks with a different regularization parameter $\alpha$: zero
(black solid curve),
small (dotted curve), large (dashed curve) and for integrated velocity (grey
curve).\label{fig4}}
\end{figure}

In order to test the efficiency of the optimal masks we compute how these masks
filter
the modes versus the target mode and compare to the spherical-harmonic masks.
 We
assume that each mode leads to a velocity signal which is a perfect spherical
harmonic
projected onto the line of sight. The degree range of modes is given by the
properties
of the detector.  The 180-bin detector (the LOI-proxy) shown in Fig. 1
is sensitive up to modes $l$=15.  The higher-degree modes
do not contribute significantly to the observed signal. Therefore, in the
global masks we restrict the mode range from $l=0$ to 15.

The results of the global and local masks are represented in the form
of a map (called hereafter amplitude map) showing the
relative amplitude of the modes identified with the doublet ($l$, $m$) for $l$
between
0 and 6 and $m$ between $-l$ and $+l$.

Figure 2 shows, as an example for a target $l$=1 and $m$=1, the way various
masks filter the unwanted modes.  On top left we used the disk-integrated
signal
which leads to the usual problem of mode blending between $m$=$\pm$ 1. It is
well-known
(Chang, 1997) that this blending  especially bias the determination of
frequency splitting.
For the spherical-harmonic masks (top right), the blending is smaller but still
there is
the same problem. Using the local optimal masks for a frequency window of
$\pm$15 $\mu$Hz we
are able to reduce below the noise level
the $m$=-1 as well as the $l$=3,6 and 9 modes which were also
interfering with the target mode.
Note that global masks (bottom, right) as expected do not
remove perfectly these modes because the number of modes to filter (around
3000)
exceeds the number of bins by a large factor.  
We give another example for a $l$=4, $m$=3
mode in
Fig. 2. Applying the local optimal masks should therefore lead to clean
power spectra around the target mode as shown in next section using the MDI
data.

\section{Optimal masks applied to MDI data}

In order to apply the optimal masks to MDI data we need in principle to take
account of
the time-dependence of the signal due to the displacement of the image on the
detector
and change of its size as the SOHO-Sun distance is changing. However, for the
low-degree
modes this effect is expected to be small, and was neglected in these studies.

The larger effect would result from the variations of the images orientation
angles
$P_0$ and $B_0$. However, in the MDI observations the $P_0$ is quite stable and
much less than 1 degree. The $B_0$ varies between -7 and + 7 degrees during
a year. However, we found that these variations lead only to non-significant
variations of our optimal masks. Therefore, we adopt the mean value, $B_{0}$=0.

As an example, Figure 4 shows overlapped two pieces of power spectra
of a $l$=1, $m=-1$ mode,  one obtained with the standard spherical
harmonic masks (grey curve) and the other one with the local optimal masks
(black curve).
In the first case, the power spectrum shows a significant leakage of the
$l$=1 $m$=1 mode which belongs to the same rotationally split multiplet.
This mode partially overlaps with the
target mode. In the optimal mask spectrum, the
contribution of this mode is reduced to the level of noise.
As previously noticed the signal-to-noise
ratio is better with the spherical-harmonic masks
but the leakage mode is strongly blended with the target one
leading to bias in the estimation of mode parameters. Modes of $l$=6 and 9
were also filtered for this case. The singular value cut-off was chosen to be
of the order of the numerical accuracy and no regularization was applied
which means optial filtering with a slightly larger noise level.

Figure 5 shows the effect of regularization for a target mode of $l$=0, where
the modes of $l$=2, 5, 8, 11 and 15 are filtered out in a window
of $\pm$15$\mu$Hz around the target mode.
For $\alpha$=0 the contiguous $l$=2 mode multiplet at 3024 $\mu$Hz
is reduced to noise level whereas as $\alpha$ increases this mode becomes
larger and the
noise level decreases. We found that in most cases the optimal masks with
$\alpha=0$
work sufficiently well.

\section{Conclusions}
Mode leakage which results in overlapping mode peaks in solar
oscillation power spectra is one of the most significant problems
in observational helioseismology. It may lead to significant errors
in the determination of mode frequencies and other properties.
We have shown that this problem can be efficiently solved by
using the optimal masks which are applied to series of solar images
and reduce the signals of solar modes in a narrow frequency
interval around a target mode to the level of noise. The optimal
mask, however, may increase the noise level. Therefore, the filtering
efficiency of the masks has to be balanced with the noise level.
This is achieved by a regularization technique. We have demonstrated
in the case of low-degree modes that the optimal mask technique
allows us efficiently isolate individual modes in rotationally
split multiplets, which are unresolved in the commonly used disk-integrated
data, and also isolate modes in the case of overlapped multiplets
of different angular degree.
The optimal masks are quite efficient for filtering out unwanted modes
as long as their number is less or of the order of the number of solar image
bins.  We used the MDI LOI-proxy data to
calculate the optimal masks. However, this method is quite
general and can be used for any spatially resolved helioseismic data.

The method has been applied for the determination of the central
low-degree frequencies by Toutain et al (1999). In a future paper,
we will present the results for rotational splitting of low-degree
modes.


\begin{references}
\reference {}
Appourchaux, T, Andersen, B., Fr\"ohlich, C., Jimenez, A., Telljohann, Udo,
Wehrli, C., 1997, Sol. Phys. 170, 27.
\reference {}
Appourchaux, T. and Gizon, L., Rabello-Soares, M.-C., 1998, A\&AS 132, 107
\reference {}
Chang, H.-Y., 1997, PhD thesis, University of Cambridge.
\reference {}
Christensen-Dalsgaard, J. 1984, in Solar Seismology From Space,
ed. R.K. Ulrich, J. Harvey, E.J. Rhodes, and J. Toomre,
Pasadena: Jet Propulsion Laboratory, 219.
\reference {}
Gough, D.O., and Latour, J. 1984, Astron. Express, 1, 9.
\reference {}
Hoeksema, J.T., Bush, R.I., Mathur, D., Morrison, M., and Scherrer, P.H., 1998,
in:
Sounding Solar and Stellar Interiors, Symp. IAU 181, eds. J. Provost, F.-X.
Schmider,
Poster Vol., Univ. Nice, p.31.
\reference {}
Kosovichev, A.G. 1986, Bull. Crimean Astrophys. Obs., 75, 19.
\reference {}
Press, W.H., Teukolsky, S.A., Vetterling, W.T., and Flannery, B.P.
Numerical Recipes, Second Edition, Cambridge University Press, 1992.
\reference {}
Scherrer,P.H., Bogart, R.S., Bush, R.I., Hoeksema, J.T., Kosovichev, A.G.,
Schou, J., Rosenberg, W., Springer, L., Tarbell, T.D., Title, A., Wolfson,
C.J., Zayer, I. and the MDI engineering team, 1995, Sol. Phys.  162, 129.
\reference {}
Toutain, T.,  Appourchaux, T., Fr\"ohlich, C.,
 Kosovichev, A. G., Nigam, R., and Scherrer, P. H., 1999, ApJ, 506, L147.
\end{references}
\end{document}